# Averaging Hypotheses in Newtonian Cosmology

Thomas Buchert[1]

*Theoretische Physik, Ludwig-Maximilians-Universität, Theresienstr. 37, D–80333 München, F.R.G.*

**Abstract.** Average properties of general inhomogeneous cosmological models are discussed in the Newtonian framework. It is shown under which circumstances the average flow reduces to a member of the standard Friedmann–Lemaître cosmologies. Possible choices of global boundary conditions of inhomogeneous cosmologies as well as consequences for the interpretation of cosmological parameters are put into perspective.

## 1. The Standard Averaging Hypothesis

The description of the Universe in terms of homogeneous solutions of Einstein's or Newton's equations for gravitationally interacting matter, respectively, is to be considered a historical remnant: Clearly, homogeneous cosmologies, in particular the family of isotropic Friedmann–Lemaître models, were motivated on observational grounds at the time of their innovation indicating a fairly homogeneous distribution of matter around us, but were not justified on theoretical grounds. While the assumption of global isotropy has survived the observational experience until today's assessment of an extreme isotropy of the microwave background, the assumption of local isotropy around every point in space (which implies global homogeneity for analytical fields) is not adequate; the homogeneous models have to be replaced by inhomogeneous cosmologies which account for the highly inhomogeneous matter distribution observed today.

   The homogeneous solutions are generally known to be linearly unstable to perturbations in the matter variables. Therefore, it cannot be expected that the standard models are reliable ones for the prediction of longtime behavior of universal dynamics. However, these models still enjoy application as "background models"; empirically used parameters like the Hubble constant, the density parameter and others are still thought to be determined by a homogeneous–isotropic solution, independent of spatial scale. Observations on scales which are *assumed to be* "statistically fair" are interpreted accordingly. This naïve description of the global dynamical properties of the inhomogeneous Universe can only be justified, if Friedmann–Lemaître cosmologies model the *average dynamics* correctly. The cosmological principle, in this context, may be considered an averaging hypothesis which asks for approval.

---

[1] E–mail: buchert@stat.physik.uni-muenchen.de

Examining present–day models of large–scale structure, either analytical approximations or numerical N–body simulations, we have to conclude that *all* models (that are actually used) are *constructed* such to obey the standard averaging hypothesis. *A priori*, however, there is no reason to believe this unless the average flow is calculated and the assumptions, which reduce the general average dynamics to a model which is a member of the Friedmann–Lemaître cosmologies, are specified. Moreover, it has been pointed out several times, especially in the recent past (Shirokov & Fisher 1963, Ellis 1984, Futamase 1989) that the inherent nonlinearity of the equations (here always referred to the equations of General Relativity) manifestly contradicts the expectation that dynamical evolution and averaging commute, i.e., that the average of an evolved matter distribution agrees with the evolution of the average. This may be true for solutions of linearized equations or, in a statistical sense, for restricted ensembles such as for models based on the assumption of homogeneous–isotropic turbulence (Olson & Sachs 1973), but will not be in general the case as soon as we describe nonlinear stages of the evolution of matter variables, which is claimed to be covered by current structure formation models.

Futamase (1989) (see also: Bildhauer 1990, Bildhauer & Futamase 1991a,b, Futamase 1995, Seljak & Hui 1995) proposes a general relativistic model to calculate the "backreaction effect" of local inhomogeneities on the global expansion law. In his model the effect is largest for strongly anisotropic gravitational collapse, inhomogeneities generally accelerate the expansion of the Universe, and they do it in a scale–dependent way. That this result has apparently not reached the community which builds and analyses models of large–scale structure may be explained by the ambiguity involved in averaging general relativistic models. Indeed, the averaging procedure is not uniquely determined, the metric enters as an additional dynamical variable (it may be averaged or, instead, deformed, as suggested by Carfora & Marzuoli 1988, Carfora et al. 1990).

Since the main reason for the existence of a nonvanishing "backreaction" (i.e., the nonlinearity of the system of equations) is also present in the Newtonian framework, we can address the problem much easier there, and we will be able to build a precise notion of how to deal with it. However, even the Newtonian limit of averaged GR dynamics has not to coincide with the averaged Newtonian theory on scales which are thought to be fully covered by a Newtonian treatment.

## 2. General Expansion Law in Newtonian Cosmology

### 2.1. Time–evolution and averaging do not commute

Let us consider a portion of the Universe $D(t)$ with volume $V(t)$. Henceforth, we concentrate on the expansion which we describe by the local expansion scalar $\theta = \nabla \cdot \vec{v}$. Introducing the scale–factor via the volume, $a_D := V^{1/3}$, we can write the spatial average of $\theta$ on the domain $D$ as (Buchert & Ehlers 1995):

$$\langle \theta \rangle_D = \frac{1}{V} \int_D d^3x \, \theta = \frac{\dot{V}}{V} = 3\frac{\dot{a}_D}{a_D} \quad . \tag{1}$$

As written in (1) the spatial average can be calculated as a simple Euclidean volume integral over the domain $D$, the main advantage of a Newtonian treatment.

The subscript $D$ indicates that the averages (as well as the scale–factor) depend on morphological properties of the spatial domain such as content, shape and connectivity.

We evaluate the evolution of the average in a tube of trajectories of fluid elements, i.e., we introduce a Lagrangian mapping $\vec{f}_t : \vec{x} = \vec{f}(\vec{X}, t)$ which sends fluid particles from their initial (Lagrangian) position $\vec{X}$ to their final (Eulerian) position $\vec{x}$. We use the Jacobian of this mapping, $d^3x = J d^3X$, $J := \det\left(\frac{\partial f_i}{\partial X_k}\right)$, to transform spatial averages to volume integrals in Lagrangian space[2]:

$$\langle \theta \rangle_D = \frac{1}{V} \int_{D(t)} d^3x \; \theta(\vec{x}, t) = \frac{1}{V} \int_{D(t_0)} d^3X \; J(\vec{X}, t) \theta(\vec{X}, t) \; . \tag{2}$$

Using the Lagrangian time–derivative, $\frac{d}{dt} := \frac{\partial}{\partial t} + \nabla \cdot$, we obtain the *nonlinear commutation rule* (Buchert & Ehlers 1995):

$$\frac{d}{dt}\langle \theta \rangle_D - \langle \frac{d}{dt} \theta \rangle_D \;=\; \langle \theta^2 \rangle_D - \langle \theta \rangle_D^2 \; . \tag{3}$$

Equation (3) shows that the evolution of the average and the average over the evolved field do not commute, their difference being given by the nonlinear fluctuation term on the r.h.s. .

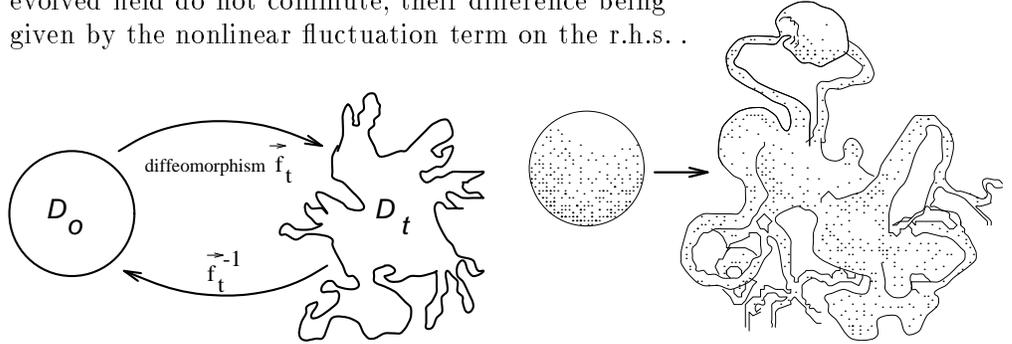

Figure 1. The mapping from Lagrangian to Eulerian space is in general not diffeomorphic; it may not even be a homeomorphism (Figure inspired by Carfora & Piotrkowska 1995).

### 2.2. The generalized Friedmann equation

Averaging Raychaudhuri's equation for the evolution of the expansion scalar, the scale factor $a_D$ is found to obey the general expansion law (Buchert & Ehlers 1995):

$$3\frac{\ddot{a}_D}{a_D} + 4\pi G \frac{M_D}{a_D^3} - \Lambda \;=\; -\mathbf{Q} \quad , \tag{4a}$$

where the source term $\mathbf{Q}$ depends on the fluctuation term in (3) and the magnitudes of rotation ($\omega$) and shear ($\sigma$) of the flow,

$$\mathbf{Q} := \frac{2}{3}\left(\langle \theta \rangle_D^2 - \langle \theta^2 \rangle_D\right) + 2\langle \sigma^2 - \omega^2 \rangle_D \quad . \tag{4b}$$

---

[2] A disclaimer is added in Figure 1.

$M_D$ denotes the total mass contained in $D$.

Eq. (4) may be rewritten as a standard Friedmann equation for the actual source term $\langle \varrho_{eff} \rangle_D$:

$$4\pi G \langle \varrho_{eff} \rangle_D := 4\pi G \langle \varrho \rangle_D + \mathbf{Q} \;, \tag{5}$$

where $\langle \varrho \rangle_D = M_D/a_D^3$ is the pure average matter density. Eq. (5) shows that, for irrotational flows (which we may consider a good assumption until the epoch of structure formation), the additional "dynamical mass" is a *positive* term which adds to the matter density, if $\sigma^2$ is larger than the fluctuation $\langle \theta \rangle_D^2 - \langle \theta^2 \rangle_D = \langle (\theta - \langle \theta \rangle_D)^2 \rangle_D \geq 0$. This suggests to add the source term $\mathbf{Q}$ to the list of *dark matter* candidates: strongly sheared inhomogeneities could "fake" an additional density which, e.g., leads to an overestimate of the density parameter.

Integrating eq. (4) with respect to time yields the generalized form of Friedmann's differential equation:

$$\frac{\dot{a}_D^2 + k}{a_D^2} - \frac{8\pi G M_D}{3 a_D^3} - \frac{\Lambda}{3} = \frac{1}{3 a_D^2} \int_{t_0}^{t} dt' \, \mathbf{Q} \, \frac{d}{dt'} a_D^2 \;. \tag{6}$$

### 2.3. Averaging globally homogeneous–isotropic universes

We now assume that a global Hubble flow exists on some large scale $A$ *and* that the expansion–factor on that scale obeys Friedmann's differential equation (Eq. (6) for $\mathbf{Q} = \mathbf{0}$); on the scale $A$ we write $a_D \equiv: a$. Splitting the velocity gradient $v_{i,j}$ into its Hubble part and a peculiar–velocity gradient, $v_{i,j} = H(t)\delta_{ij} + u_{i,j}$, where $H(t) = \frac{\dot{a}}{a}$, we obtain:

$$\theta = 3H + \nabla \cdot \vec{u} \;. \tag{7}$$

After averaging and using Gauß's theorem, the last equation leads to a relation between the Hubble function $H(t)$, the "effective Hubble function" $H_D(t) := \frac{\dot{a}_D}{a_D}$, and the peculiar–velocity field $\vec{u}(\vec{x}, t)$:

$$H(t) = H_D(t) - \frac{1}{3} a_D^{-3} \int_{\partial D(t)} d\vec{S} \cdot \vec{u} \;. \tag{8}$$

$H_D$ may be interpreted as that Hubble function which is inferred from the (possibly anisotropic and rotational) dynamics of the spatial domain $D$. (This interpretation is possible if statistical averages of many such spatial domains are considered, but, at present, we only measure one member of such an ensemble.)

Accordingly, the source term $\mathbf{Q}$ can be split into its Hubble part and deviations thereof and transformed into surface integrals over the boundary $\partial D(t)$ (Buchert & Ehlers 1995):

$$3\frac{\ddot{a}_D}{a_D} + 4\pi G \frac{M_D}{a_D^3} - \Lambda = -\frac{2}{3} \left( a_D^{-3} \int_{\partial D} d\vec{S} \cdot \vec{u} \right)^2 + a_D^{-3} \int_{\partial D} d\vec{S} \cdot (\vec{u} \nabla \cdot \vec{u} - \vec{u} \cdot \nabla \vec{u}) \;. \tag{9}$$

### 2.4. Fair samples and cosmological parameters

In view of Eqs. (6), (8) and (9) the measured average characteristics like Hubble's constant, the density parameter, and others will, in general, show a scale dependence. The Hubble constant may be large on small scales where the effect of **Q** dominates, but small on larger scales; the density parameter shows the same tendency in conformity with the quantitative calculations by Bildhauer & Futamase (1991a,b). Consequently, the age of the Universe is likely to be underestimated on scales which we currently access observationally. Thus, looking at the presently discussed observational values of these parameters and the problems of large–scale structure models, this effect works in the right direction: Models for large–scale structure favor a small Hubble constant, possibly as small as 30 km/sMpc as championed by Bartlett et al. (1995), observations of the local Universe point instead to a value of about 60 − 70 km/sMpc (Ellis et al. 1995), which has to be identified with the "effective Hubble function" as inferred from one spatial domain (our environment).

Let us have a closer look at the "backreaction terms" in Eq. (9). Both terms are entirely neglected in the traditional approach, because one assumes that $\int_D d^3x \, \nabla \cdot \vec{u} = 0$ on any postulated "fair sample" of the Universe (even if the sample is of the order of 100 Mpc shallow); also the other fluctuation term is neglected by construction due to the assumption that the matter variables average to zero on their assumed Friedmann backgrounds. Concerning the first term it is interesting to note that reconstruction models which try to recover the density distribution from measurements of peculiar–velocities are determined only up to a constant offset factor (the mean density contrast $\langle \delta \rangle_D$ on the observed scale, which is related to the square root of the first term in (9) using linear perturbation theory, $\delta \propto -\nabla \cdot \vec{u}$; see, e.g., Dekel 1994).

The qualitative considerations above, however, indicate that estimates on intermediate and small scales will give the worst results for the mean characteristics of the Universe. The source term **Q** will be most effective in Λ–dominated universes where fluctuations grow fast. To establish this effect quantitatively we could run large (Gpc–)simulations and look at sampling variations in the quantity $[\frac{1}{3}(\langle\theta\rangle_D^2 - \langle\theta^2\rangle_D) + \langle\sigma^2 - \omega^2\rangle_D]/2\pi G \langle\varrho\rangle_D$. Previous estimates of fluctuations in the cosmological parameters were based on spherical symmetric models ($\sigma = 0$) which can only account for the fluctuation term for $\theta$ (e.g., Suto et al. 1995). Since structures form in a highly anisotropic way, the shear term will play an important, if not dominant role. It should be emphasized that not only amplitude effects matter which are mirrored in the scale–dependence of the r.m.s. density contrast or the power spectrum, but phase correlations can produce large (albeit low–amplitude) structures (Beacom et al. 1991, Buchert & Martínez 1993). As long as we do not cover scales which are considerably larger than the extent of these structures and as long as the phases are not uncorrelated, we can neither expect the "backreaction effect" to vanish, nor can we speak about a "fair sample" of the Universe. Generally speaking, since there is little parameter space left for the standard model (even in view of a conservative interpretation of observational and theoretical constraints, Bagla et al. 1995, Ellis et al. 1995), there *is* a problem which could be due to the ignorance of "backreaction effects".

## 3. Global Interpretations and Alternative Views

### 3.1. Toroidal cosmologies

We may assume that, on the largest scale $A$, the cosmology is topologically *closed*, i.e. space sections are compact without boundary. The simplest example of such a space form is provided by the familiar hypertorus. As a consequence of this assumption the surface integrals in eq. (8) and (9) vanish and $H_D(t) = H(t)$ on the scale $A$. (Note that the existence of a global Hubble flow represents a restriction of generality in the case of spatially compact models, because there might exist a global vorticity and/or shear flow on the scale $A$; see Ehlers & Buchert 1995.) Since toroidal cosmologies are realized if the perturbation fields are periodic on the box size of a simulation, we conclude that, technically, current structure formation models obey the standard averaging hypothesis for globally homogeneous–isotropic universes, i.e., they average out to the standard background models, which has been assumed so far without proof. However, the scale $A$ is usually set *ad hoc*; the present analysis shows that it will be only meaningful to "close the topology" on a scale where the "backreaction terms" discussed have negligible influence, unless we deal with "small universes" (Ellis 1971) having literally a closed topology.

We have to be careful by extending this picture to General Relativity, e.g., for $\Omega < 1$ universes the space sections have constant negative curvature and cannot be compactified to a hypertorus. (For details on possible space forms see, e.g., Ellis 1971, Lachièze–Rey & Luminet 1995 and ref. therein.)

### 3.2. Infinite and topologically open cosmologies

Although compact and unbounded cosmologies are attractive, if not favourable, it is possible that there doesn't exist *any* scale on which the "backreaction" vanishes, and there may be no Hubble flow which fixes a standard frame of reference. However, we believe that the source term **Q** has smaller values on larger scales which may entitle us to "approximate" the Universe by standard models on the largest scales. However, the bare existence of **Q** indicates the presence of an intimate relationship between the evolution of inhomogeneities and the morphology of patches of smoothed–out matter distributions on *every* scale. To ignore this term entails an attitude towards "closing the eyes" in front of a possibly exciting development of inhomogeneous models of the Universe. A reinvestigation of globally hierarchical cosmologies is only possible *with* the source term **Q**.

### 3.3. Scaling, coarse–graining and renormalization

Armed with tools for a scale–dependent average dynamics, we might accomplish a fully scale–dependent dynamical description of the Universe by advancing renormalization group techniques. Having emerged from different branches of physics, these techniques might initialize the future of inhomogeneous cosmology, calling for the establishment of the "coarse graining" idea, which could shed light on general scaling properties of both gravitational dynamics and statistical characteristics of the matter distribution. The urgency of such an approach has already invoked various efforts (e.g., Carfora & Piotrkowska 1995, Piotrkowska 1995, Pérez–Mercader et al. 1995).

**Acknowledgments.** I would like to thank Jürgen Ehlers, Bernard Jones, Kamilla Piotrkowska, Enn Saar, José–Luis Sanz and Herbert Wagner for valuable discussions. This work was supported by the "Sonderforschungsbereich **375** für Astro–Teilchenphysik der Deutschen Forschungsgemeinschaft".

**References**

Bagla, J.S., Padmanabhan, T. & Narlikar, J.V. 1995, IUCAA preprint, astro-ph/9511102

Bartlett, J.G., Blanchard, A., Silk, J. & Turner, M.S. 1995, Science, 267, 980

Beacom, J.F., Dominik, K.G., Melott, A.L., Perkins, S.P. & Shandarin, S.F. 1991, ApJ, 372, 351

Bildhauer, S. 1990, Prog.Theor.Phys., 84, 444

Bildhauer, S. & Futamase, T. 1991a, MNRAS, 249, 126

Bildhauer, S. & Futamase, T. 1991b, G.R.G., 23, 1251

Buchert, T., Martínez, V.J. 1993, ApJ, 411, 485

Buchert, T., & Ehlers, J. 1995, MNRAS, submitted

Carfora, M., Marzuoli, A. 1988, Class.Quant.Grav., 5, 659

Carfora, M., Isenberg, J. & Jackson, M. 1990, J.Diff.Geom., 31, 249

Carfora, M. & Piotrkowska, K. 1995, Phys.Rev.D, 52, 4393

Dekel, A. 1994, Annu.Rev.Astron.Astrophys., 32, 371

Ehlers, J. & Buchert, T. 1995, in preparation

Ellis, G.F.R. 1984, in Proc. 10th international conference on General Relativity and Gravitation, B. Bertotti et al., Dordrecht: Reidel, 215-288

Ellis, G.F.R. 1971, G.R.G., 2, 7

Ellis, G.F.R., Ehlers, J., Börner, G., Buchert, T., Hogan, C.J., Kirshner, R.P., Press, W.H., Raffelt, G., Thielemann, F–K. & Van den Bergh, S. 1995, in Dahlem Workshop Report ES19 "The Evolution of the Universe", S. Gottlöber, G. Börner, Chichester: Wiley, in press

Futamase, T. 1989, MNRAS, 237, 187

Futamase, T. 1995, Phys.Rev.D., in press

Lachièze–Rey, M. & Luminet, J.–P. 1995, Phys.Rep., 254, 135

Olson, D.W. & Sachs, R.K. 1973, ApJ, 185, 91

Pérez–Mercader, J., Goldman, T., Hochberg, D. & Laflamme, R. 1995, Phys.Rev.Lett., submitted, astro-ph/9506127

Piotrkowska, K. 1995, U.C.T. preprint, gr-qc/9508047

Seljak, U. & Hui, L. 1995, Proc. "Clusters, Lensing and the Future of the Universe", College Park, Maryland, in press

Shirokov, M.F. & Fisher, I.Z. 1963, Sov.Astron., 6, 699

Suto, Y., Suginohara, T. & Inagaki, Y. 1995, Prog.Theor.Phys., 93, 839